\documentclass[10pt,a4paper]{article}
\pdfoutput=1
\usepackage[dvipsnames]{xcolor}
\usepackage{amsmath,fourier,amssymb,amsthm,graphicx,epstopdf,tikz,hyperref,
	color,cases,float,standalone}
\usetikzlibrary{shapes,snakes}
\usetikzlibrary{calc,backgrounds,patterns,arrows.meta}
\usepackage{chngcntr}
\counterwithout{figure}{section}
\usepackage[T1]{fontenc}
\allowdisplaybreaks
\numberwithin{equation}{section}

\newtheorem{innercustomthm}{Theorem}
\newenvironment{customthm}[1]
{\renewcommand\theinnercustomthm{#1}\innercustomthm}
{\endinnercustomthm}

\theoremstyle{plain}
\newtheorem {hypo}{\bf\hspace{-\parindent}Hypothesis}[section]

\newtheorem {prop}[hypo]{Proposition}
\newtheorem {lemma}[hypo]{Lemma}

\newtheorem {cor}[hypo]{Corollary}

\theoremstyle{remark}

\newcommand{\pf}{\begin{bpf}}

	\newcommand{\pfms}{\begin{bpfms}}
		\newcommand{\epf}{\end{bpf}\hfill$\square$\vspace{0.1cm}}
	\newcommand{\epfms}{\end{bpfms}\hfill$\square$\\ }
\newcommand\ben{\begin{equation*}}
\newcommand\ebn{\end{equation*}}
\newcommand\beq{\begin{equation}}
\newcommand\eeq{\end{equation}}
\newcommand\ds{\displaystyle}
\newcommand\lb{\left(}
\newcommand\rb{\right)} 

\newcommand\Cb{\mathbb{C}} 

\newcommand\Pb{\mathbb{P}}

\begin{document}

	\LARGE
	\noindent
	\textbf{Perturbative connection formulas for Heun equations}
	\normalsize
	\vspace{1cm}\\
	\textit{ 
		O. Lisovyy$\,^{a}$\footnote{lisovyi@lmpt.univ-tours.fr}, A. Naidiuk$\,^{a}$\footnote{andrii.naidiuk@lmpt.univ-tours.fr}}
	\vspace{0.2cm}\\
	$^a$  Institut Denis-Poisson, Universit\'e de Tours, CNRS, Parc de Grandmont,
	37200 Tours, France
	
	\begin{abstract}
	Connection formulas relating Frobenius solutions of linear ODEs at different Fuchsian singular points can be expressed in terms of the large order asymptotics of the corresponding power series. We demonstrate that for the usual, confluent and reduced confluent Heun equation, the series expansion of the relevant asymptotic amplitude in a suitable parameter can be systematically computed to arbitrary order. This allows to check a recent conjecture of Bonelli-Iossa-Panea Lichtig-Tanzini expressing the Heun connection matrix in terms of quasiclassical Virasoro conformal blocks.	
	\end{abstract}

    \section{Introduction}
    Heun's differential equation was introduced in 1889 \cite{Heun} as the general linear 2nd order ODE with four Fuchsian singular points. Being the simplest generalization of the Gauss hypergeometric equation, it is at the same time highly nontrivial  because of a qualitatively novel feature --- the presence of accessory parameters. A long list of applications of Heun's equation and its confluent versions includes such diverse topics as e.g. the theory of conformal mappings, Painlevé functions, black hole scattering and quantum optics; see \cite{Ronveaux,Hortacsu} for an extensive bibliography.

    It was discovered by Zamolodchikov \cite{Zamo86} that Heun accessory parameter function is directly related to the quasiclassical 4-point Virasoro conformal block. The Heun equation appears in this setting \cite{T10,LLNZ} as a limit of the BPZ decoupling equations \cite{BPZ} for conformal blocks with degenerate insertions. Under the AGT correspondence \cite{AGT}, the quasiclassical limit of Liouville CFT corresponds to the Nekrasov-Shatashvili limit \cite{NS} of the corresponding 4D $\mathcal N=2$ supersymmetric $\mathrm{SU}(2)$ gauge theories. This has brought a new  surge of interest in the study of equations of Heun type and their applications  in the past decade, see e.g. \cite{AGH,BGG,CCC,CCN,LeNo,LN,NC,PP14,PP17}.
    
    A remarkable further step was recently made in \cite{BILT21,BILT22} where it was realized that (a class of) connection problems for Heun equations are also solvable in terms of quasiclassical conformal blocks. The main idea behind this statement is the calculation of quantum (operator-valued) monodromy of conformal blocks with respect to positions of degenerate fields using the elementary fusion and braiding transformations\footnote{The approach of  loc.~cit. uses in addition the DOZZ formula for Liouville structure constants. However, it is not strictly necessary: the locality of the elementary fusion/braiding operations is already sufficient to arrive to the same result, as discussed in \cite{Lis,CFMP} and Section~\ref{sectionCFT} of the present manuscript.}. The procedure is closely related to the study of Verlinde loop operators in \cite{AGGTV,DGOT} and it has already been used in \cite{ILT} to express solutions of Fuchsian systems and associated isomonodromic tau functions via ${c=1}$ conformal blocks. The Heun connection problem similarly emerges in the analysis of the $c\to\infty$  limit of quantum monodromy.
     In practical terms, the relation to CFT allows to compute explicitly a perturbative expansion of the Heun connection matrix in a suitable parameter that can be efficiently used in applications \cite{CFMP,DGILZ}.
   
    The aim of the present work is to show  that equivalent series expansions of the Heun connection coefficients can be derived in a rigorous way, without relying on CFT intuition. First terms of these expansions and their counterparts for Mathieu equation can indeed be computed using the procedure of \cite[Section 7]{HK} and \cite[Subsection 5.4]{HRS} of perturbative construction of global solutions. In principle, one may try to extend it order by order to compute subleading corrections. However, to obtain an all-order result in a compact form, we employ instead a different approach --- a general result of Schäfke and Schmidt \cite{SS} which expresses the connection matrix relating two Frobenius bases of solutions of a linear 2nd order ODE at different Fuchsian singular points in terms of the large-order asymptotics of their expansion coefficients.
    
    Our main results are summarized as follows.
   \begin{customthm}{A}\label{thmA}
    Let us write the normal form of the reduced confluent Heun equation as
   	\beq
   	\left[\frac{d^2}{dz^2}+\frac{\frac14-\theta_0^2}{z^2}+\frac{\frac14-\theta_1^2}{\lb z-1\rb^2}+\frac{\theta_0^2+\theta_1^2-\omega^2-\tfrac14-\lambda z}{z\lb z-1\rb}\right]\psi\lb z\rb=0,
   	\eeq
    with $\theta_0,\theta_1\notin\mathbb Z/2$.  Denote by $\psi^{[0]}_{\pm}\lb z\rb$, $\psi^{[1]}_{\pm}\lb z\rb$ its Frobenius solutions at $z=0,1$   	normalized as
   	\begin{subequations}\label{norfrob}
   	\begin{alignat}{3}
   	\psi^{[0]}_{\pm}\lb z\rb=&\,z^{\frac12\mp\theta_0}\left[1+\sum_{k=1}^\infty \psi^{[0]}_{\pm,k} z^k\right]&\qquad \text{as }z\to0^+, \\ 
   	\psi^{[1]}_{\pm}\lb z\rb=&\,\lb 1-z\rb^{\frac12\mp\theta_1}\left[1+\sum_{k=1}^\infty \psi^{[1]}_{\pm,k} \lb z-1\rb^k\right]&\qquad \text{as }z\to1^-.
   	\end{alignat}
   	\end{subequations}
    The connection between the two Frobenius bases is given by
    \beq\label{connnn}
    \psi^{[0]}_{\epsilon}\lb z\rb=\sum_{\epsilon'=\pm}\;\mathsf{C}\lb \epsilon\theta_0,\epsilon'\theta_1\rb
    \psi^{[1]}_{\epsilon'}\lb z\rb,\qquad\qquad \epsilon=\pm,
    \eeq
    where the function $\mathsf C\lb\theta_0,\theta_1\rb$ admits the following representation in terms of continued fractions:
    \beq\label{CoRCHE}
    \mathsf C\lb\theta_0,\theta_1\rb=\frac{\Gamma\lb 1-2  \theta_0\rb\Gamma\lb2\theta_1\rb}{\Gamma\lb\tfrac12+\theta_1-\theta_0+\omega\rb \Gamma\lb\tfrac12+\theta_1-\theta_0-\omega\rb}\,\exp \sum_{k=1}^\infty\ln\lb 1-\frac{\lambda\beta_k}{1-\frac{\lambda\beta_{k+1}}{1-\ldots}}\rb,
    \eeq
    with
      \beq\label{betaRCHE}
    \beta_k=\frac{k\lb k-2\theta_0\rb}{\lb \lb k+\frac12-\theta_0+\theta_1\rb^2-\omega^2\rb
    	\lb \lb k-\frac12-\theta_0+\theta_1\rb^2-\omega^2\rb}.
    \eeq
   	\end{customthm}
   \begin{customthm}{B}\label{thmB}
   	Consider the normal form of the Heun equation,
   	\beq
   	\left[\frac{d^2}{dz^2}+\frac{\frac14-\theta_0^2}{z^2}+\frac{\frac14-\theta_1^2}{\lb z-1\rb^2}+\frac{\frac14-\theta_t^2}{\lb z-t\rb^2}+\frac{\theta_0^2+\theta_1^2+\theta_t^2-\theta_\infty^2-\tfrac12}{z\lb z-1\rb}+\frac{\lb t-1\rb \lb \omega^2+\theta_t^2-\theta_\infty^2-\tfrac14\rb}{z\lb z-1\rb\lb z-t\rb}\right]\psi\lb z\rb=0,
   	\eeq
    and assume that $|t|>1$ and $\theta_0,\theta_1\notin\mathbb Z/2$. Let $\psi^{[0]}_{\pm}\lb z\rb$, $\psi^{[1]}_{\pm}\lb z\rb$ denote its Frobenius solutions normalized as in \eqref{norfrob}. The relation between the two bases is given by \eqref{connnn}, where the  function $\mathsf C\lb\theta_0,\theta_1\rb$ can be written as
     \beq\label{CoHE}
    \mathsf C\lb\theta_0,\theta_1\rb=\frac{\Gamma\lb 1-2  \theta_0\rb\Gamma\lb2\theta_1\rb\, \lb1-\lambda\rb^{-\frac12-\theta_t}}{\Gamma\lb\tfrac12+\theta_1-\theta_0+\omega\rb \Gamma\lb\tfrac12+\theta_1-\theta_0-\omega\rb}\,\exp \sum_{k=1}^\infty\ln\lb1-\lambda\alpha_{k-1}-\frac{\lambda\beta_k}{1-\lambda\alpha_{k}-\frac{\lambda\beta_{k+1}}{1-\ldots}}\rb,
    \eeq
    with $\lambda=\frac1t$ and
     \begin{subequations}\label{alphabetaHE}
    	\begin{align}
    	\label{alphaHE}
    	\alpha_k=&\, -\frac{\lb k+\frac12-\theta_0-\theta_t\rb^2-\theta_0^2-\theta_\infty^2+\omega^2}{\lb k+\frac12-\theta_0+\theta_1\rb^2-\omega^2},\\
    	\label{betaHE}
    	\beta_k=&\,\frac{k\lb k-2\theta_0\rb\lb
    		\lb k-\theta_0+\theta_1-\theta_t\rb^2-\theta_\infty^2\rb}{\lb \lb k+\frac12-\theta_0+\theta_1\rb^2-\omega^2\rb
    		\lb \lb k-\frac12-\theta_0+\theta_1\rb^2-\omega^2\rb}.
    	\end{align} 	
    \end{subequations}
   \end{customthm}

    Explicit perturbative solution of the connection problem for small $\lambda$ is obtained by truncating the infinite fractions in \eqref{CoRCHE} and \eqref{CoHE} at the desired order, expanding the result in powers of $\lambda$ and algorithmically computing the resulting infinite sums in terms of polygamma functions. The results match the coefficients of the perturbative series predicted by CFT considerations from \cite{BILT21,BILT22}. 

   The paper is organized as follows. In Section~\ref{sectionCFT}, we recall the arguments leading to the conjectural solution \eqref{triestef} of the Heun connection problem in terms of quasiclassical conformal blocks. Section~\ref{sectionSS} discusses the Schäfke-Schmidt connection formula (Corollary~\ref{CorSS}). The proofs of Theorems~\ref{thmA} and~\ref{thmB} as well as the corresponding result for the confluent Heun equation are given in Section~\ref{sectionHeun}. We conclude with a brief discussion of open questions.

    \vspace{0.2cm}
    
    {\small
    \noindent\textbf{Acknowledgements}. The authors are grateful to G. Bonelli, C. Iossa, D. Panea Lichtig, A. Tanzini, B. Carneiro da Cunha and V. P. Gusynin for stimulating discussions.  }

	\section{CFT derivation of  Trieste formula}\label{sectionCFT}
	\subsection{Fusion transformations of conformal blocks}
	Let us choose $n\ge 3$ distinct points $\mathbf{t}=\left\{t_k\right\}_{k=0,\ldots,n-1} $ on the Riemann sphere $\mathbb C\mathbb P^1$. We fix from the outset ${t_0=0}$, $t_{1}=1$, $t_{n-1}=\infty$ and assume that $|t_1|<|t_2|<\ldots<|t_{n-2}|$. Spherical Virasoro conformal block $\mathcal F\lb\mathbf{t}\rb$ is a multivariate series that can be assigned to any trivalent tree with  $n$ external vertices labeled by $\mathbf t$ whose every edge~$e$ is equipped with a weight  $\Delta_e\in\mathbb C$. For example, the tree
	 
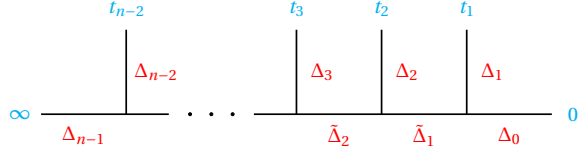
\begin{center} 
	 	\begin{tikzpicture}[scale=0.7]

\draw[thick] (0,0) -- (3,0);
\draw[thick] (2,0) -- (2,2);
\filldraw[black] (3.5,0) circle (1pt);
\filldraw[black] (4,0) circle (1pt);
\filldraw[black] (4.5,0) circle (1pt);
\draw[thick] (6,0) -- (6,2);
\draw[thick] (8,0) -- (8,2);
\draw[thick] (10,0) -- (10,2);
\draw[thick] (5,0) -- (12,0);

\node[cyan] () at (-0.5, 0){$\infty$};
\node[cyan] () at (2, 2.5){$t_{n-2}$};
\node[cyan] () at (6, 2.5){$t_3$};
\node[cyan] () at (8, 2.5){$t_2$};
\node[cyan] () at (10, 2.5){$t_1$};
\node[cyan] () at (12.5, 0){$0$};

\node[red] () at (1, -0.5){$\Delta_{n-1}$};
\node[red] () at (7, -0.5){$\tilde{\Delta}_{2}$};
\node[red] () at (9, -0.5){$\tilde{\Delta}_{1}$};
\node[red] () at (11, -0.5){${\Delta}_{0}$};

\node[red] () at (2.7, 1){$\Delta_{n-2}$};
\node[red] () at (6.6, 1){$\Delta_{3}$};
\node[red] () at (8.6, 1){$\Delta_{2}$};
\node[red] () at (10.6, 1){$\Delta_{1}$};

\end{tikzpicture}
	\end{center} 
    is represented by a series of the form
    \beq\label{auxe01}
    \begin{gathered}
    \mathcal F\lb\mathbf{t}\rb=\mathcal F_{\text{3pt}}\lb\mathbf{t}\rb \hat{\mathcal F}\lb\mathbf{t}\rb,\\ \mathcal F_{\text{3pt}}\lb\mathbf{t}\rb=\prod_{\ell=1}^{n-2}
    t_\ell^{\tilde{\Delta}_{\ell}-\tilde\Delta_{\ell-1}-\Delta_{\ell}},\qquad \hat{\mathcal F}\lb\mathbf{t}\rb=\sum_{\mathbf{k}\in \mathbb{N}^{n-3}}
    \mathcal F_{\mathbf{k}}\;
    \lb\frac{t_1}{t_2}\rb^{k_1}\lb\frac{t_2}{t_3}\rb^{k_2}\ldots
    \lb\frac{t_{n-3}}{t_{n-2}}\rb^{k_{n-3}},
    \end{gathered}
    \eeq
    where $\tilde\Delta_{0}=\Delta_0$, $\tilde\Delta_{n-2}=\Delta_{n-1}$. The coefficients $\mathcal F_{\mathbf k}$ are rational functions of the conformal weights $\boldsymbol{\Delta}$, $\boldsymbol{\tilde\Delta}$ and the Virasoro central charge $c$. They are uniquely determined by the Virasoro commutation relations and normalization ${\mathcal F_{\mathbf 0}=1}$. It will be convenient for us to parameterize the central charge and conformal weights as
    \beq
    c=1+6\lb b+b^{-1}\rb^2,\qquad \Delta=\tfrac14\lb b+b^{-1}\rb^2-p^2,
    \eeq
     Conformal block series are believed to be convergent and analytically continuable to the universal cover of the configuration space $\mathsf{Conf}_n\lb\Cb\Pb^1\rb$. Different trees with the same external dimensions give rise to conformal blocks related by sequences of elementary fusion and braiding transformations.
     
     Of special interest for us are conformal blocks involving an additional degenerate field $\Phi_{(1,2)}\lb z\rb$. Such a field has Liouville momentum $p_{(1,2)}=-b- {b^{-1}}/2$ and will be represented  by a dashed line in the diagrams below. Two important properties of degenerate fields are
     \begin{itemize}
     	\item Simple \textit{fusion rules}: the OPE of $\Phi_{(1,2)}$ with a generic Virasoro primary with momentum $p$ contains only two conformal families with momenta $p_{\pm}=p\pm\frac b2$. 
     	\item BPZ \textit{decoupling equations}: these are linear homogeneous PDEs  satisfied by the relevant conformal blocks. For $t_0$, $t_{1}$, $t_{n-1}$ fixed at 0, 1 and $\infty$, they have the form
     	$\mathcal D_{\mathrm{BPZ}}\mathcal F\lb\mathbf{t},z\rb=0$, with
    \beq
    \mathcal D_{\mathrm{BPZ}}=\frac{1}{b^2}\frac{\partial^2}{\partial z^2} -\lb\frac1z+\frac1{z-1}\rb\frac{\partial}{\partial z}+\sum_{k=2}^{n-2}\frac{t_k\lb t_k-1\rb}{z\lb z-1\rb\lb z-t_k\rb}\frac{\partial}{\partial t_k}+\sum_{k=0}^{n-2}\frac{\Delta_k}{\lb z-t_k\rb^2}+\frac{\Delta_{n-1}-\Delta_{(1,2)}-\sum_{k=0}^{n-2}\Delta_k}{z\lb z-1\rb}.
    \eeq 	
     \end{itemize}
 
    In the case of three generic primaries and one degenerate field, the BPZ equation becomes an ODE equivalent to Gauss hypergeometric equation. The corresponding 3+1 point conformal blocks provide bases of its Frobenius solutions at different Fuchsian singular points $0$, $1$, $\infty$, depending on the choice of the generic edge the degenerate field is attached to. In particular,
    \beq
    \vcenter{\hbox{\begin{tikzpicture}[scale=0.7]

\coordinate (o) at (0,0);
\coordinate (f) at ($(o)+(-30:1.3)$);

\draw[thick] (o) -- (90:2);
\draw[thick] (o) -- (-30:2);
\draw[thick] (o) -- (210:2);
\draw[thick, dashed] (f) -- +(60:1.5);

\node[cyan] () at ($(o)+(90:2.5)$){$1$};
\node[cyan] () at ($(f)+(60:1.7)$){$z$};
\node[cyan] () at ($(o)+(-30:2.5)$){$0$};
\node[cyan] () at ($(o)+(210:2.5)$){$\infty$};

\node[red] () at ($(o)+(90:1)+(180:0.3)$){$p_1$};
\node[red, scale=0.8] () at ($(o)+(-30:0.5)+(60:0.5)$){$p_0\pm\frac{b}{2}$};
\node[red] () at ($(f)+(60:1.2)+(-30:0.7)$){$p_{(1,2)}$};
\node[red] () at ($(o)+(-30:1.8)+(60:0.3)$){$p_0$};
\node[red] () at ($(o)+(210:1)+(120:0.3)$){$p_\infty$};

\end{tikzpicture} 
    }}=\mathcal F_{\mp p_{0},p_1,p_\infty}\lb z\rb,\qquad
    \vcenter{\hbox{\begin{tikzpicture}[scale=0.7]

\coordinate (o) at (0,0);
\coordinate (f) at ($(o)+(90:1.3)$);

\draw[thick] (o) -- (90:2);
\draw[thick] (o) -- (-30:2);
\draw[thick] (o) -- (210:2);
\draw[thick, dashed] (f) -- +(0:1.5);

\node[cyan] () at ($(o)+(90:2.5)$){$1$};
\node[cyan] () at ($(f)+(0:1.7)$){$z$};
\node[cyan] () at ($(o)+(-30:2.5)$){$0$};
\node[cyan] () at ($(o)+(210:2.5)$){$\infty$};

\node[red] () at ($(o)+(90:1.5)+(180:0.3)$){$p_1$};
\node[red, scale=0.8] () at ($(o)+(90:0.5)+(0:0.7)$){$p_1\pm\frac{b}{2}$};
\node[red] () at ($(f)+(0:1)+(90:0.3)$){$p_{(1,2)}$};
\node[red] () at ($(o)+(-30:1.2)+(60:0.4)$){$p_0$};
\node[red] () at ($(o)+(210:1)+(120:0.3)$){$p_\infty$};

\end{tikzpicture}
    	 }}=\mathcal F_{\mp p_{1},p_0,p_\infty}\lb 1-z\rb,
    \eeq
    where
    \beq
    \mathcal F_{p_{0},p_1,p_\infty}\lb z\rb=z^{\frac{1+b^2}{2}+ b p_0}\lb 1-z\rb^{\frac{1+b^2}{2}+ b p_1}{}_2F_1\biggl[\begin{array}{c}
    	\tfrac12+ b\lb p_1 +p_\infty+ p_0\rb,\tfrac12+ b\lb p_1 -p_\infty+ p_0\rb\\
    	1+ 2bp_0
    \end{array};z\,\biggr].
    \eeq 
    The classical connection formulas for $_2F_1$ hypergeometric functions can be interpreted as the fusion transformation for 3+1 point conformal blocks,
    \beq
    \vcenter{\hbox{\begin{tikzpicture}[scale=0.7]

\coordinate (o) at (0,0);
\coordinate (f) at ($(o)+(-30:1.3)$);

\draw[thick] (o) -- (90:2);
\draw[thick] (o) -- (-30:2);
\draw[thick] (o) -- (210:2);
\draw[thick, dashed] (f) -- +(60:1.5);

\node[cyan] () at ($(o)+(90:2.5)$){$1$};
\node[cyan] () at ($(f)+(60:1.7)$){$z$};
\node[cyan] () at ($(o)+(-30:2.5)$){$0$};
\node[cyan] () at ($(o)+(210:2.5)$){$\infty$};

\node[red] () at ($(o)+(90:1)+(180:0.3)$){$p_1$};
\node[red, scale=0.8] () at ($(o)+(-30:0.5)+(60:0.5)$){$p_0+\frac{\epsilon b}{2}$};
\node[red] () at ($(f)+(60:1.2)+(-30:0.7)$){$p_{(1,2)}$};
\node[red] () at ($(o)+(-30:1.8)+(60:0.3)$){$p_0$};
\node[red] () at ($(o)+(210:1)+(120:0.3)$){$p_\infty$};

\end{tikzpicture} 
    }}\quad=\quad\sum_{\epsilon'}\; \mathsf{F}_{\epsilon\epsilon'}\lb p_0,p_1,p_\infty\rb \vcenter{\hbox{ \begin{tikzpicture}[scale=0.7]

\coordinate (o) at (0,0);
\coordinate (f) at ($(o)+(90:1.3)$);

\draw[thick] (o) -- (90:2);
\draw[thick] (o) -- (-30:2);
\draw[thick] (o) -- (210:2);
\draw[thick, dashed] (f) -- +(0:1.5);

\node[cyan] () at ($(o)+(90:2.5)$){$1$};
\node[cyan] () at ($(f)+(0:1.7)$){$z$};
\node[cyan] () at ($(o)+(-30:2.5)$){$0$};
\node[cyan] () at ($(o)+(210:2.5)$){$\infty$};

\node[red] () at ($(o)+(90:1.5)+(180:0.3)$){$p_1$};
\node[red, scale=0.8] () at ($(o)+(90:0.5)+(0:0.7)$){$p_1+\frac{\epsilon' b}{2}$};
\node[red] () at ($(f)+(0:1)+(90:0.3)$){$p_{(1,2)}$};
\node[red] () at ($(o)+(-30:1.2)+(60:0.4)$){$p_0$};
\node[red] () at ($(o)+(210:1)+(120:0.3)$){$p_\infty$};

\end{tikzpicture}
}},\qquad \qquad \epsilon,\epsilon'=\pm.
    \eeq
    The elements of the fusion matrix $\mathsf{F}_{\epsilon\epsilon'}\lb p_0,p_1,p_\infty\rb=\mathsf{F}\lb \epsilon p_0,\epsilon' p_1,p_\infty\rb$ do not depend on $z$ and are explicitly expressed in terms of gamma functions:
    \beq\label{fusionmatr}
    \mathsf F\lb p_0,p_1,p_\infty\rb=\frac{\Gamma\lb 1-2b  p_0\rb\Gamma\lb2bp_1\rb}{\Gamma\lb\tfrac12+b\lb p_1-p_0+p_\infty\rb\rb \Gamma\lb\tfrac12+b\lb p_1-p_0-p_\infty\rb\rb}.
    \eeq
    Locality of the fusion transformations implies that 
     \beq\label{exactfusion}
    \vcenter{\hbox{\begin{tikzpicture}[scale=0.7]

\coordinate (o) at (0,0);
\coordinate (f) at ($(o)+(-60:1.3)$);
\coordinate (b) at ($(o)+(180:2)$);

\draw[thick] (o) -- (60:2);
\draw[thick] (o) -- (-60:2);
\draw[thick] (o) -- (180:2);
\draw[thick, dashed] (f) -- +(30:1.5);

\filldraw[thick, pattern = north east lines, pattern color = gray, rotate=45] (b) rectangle +(-2,2);

\node[cyan] () at ($(o)+(60:2.5)$){$1$};
\node[cyan] () at ($(f)+(30:1.7)$){$z$};
\node[cyan] () at ($(o)+(-60:2.5)$){$0$};

\node[red] () at ($(o)+(60:1)+(150:0.3)$){$p_1$};
\node[red, scale=0.8] () at ($(o)+(-60:0.5)+(30:0.7)$){$p_0+\frac{\epsilon b}{2}$};
\node[red] () at ($(f)+(30:1.2)+(-60:0.7)$){$p_{(1,2)}$};
\node[red] () at ($(o)+(-60:1.8)+(-150:0.3)$){$p_0$};
\node[red] () at ($(o)+(180:1)+(-90:0.4)$){$p_\sigma$};

\end{tikzpicture}
    	}}\quad=\quad\sum_{\epsilon'}\; \mathsf{F}_{\epsilon\epsilon'}\lb p_0,p_1,p_\sigma\rb \vcenter{\hbox{ \begin{tikzpicture}[scale=0.7]

\coordinate (o) at (0,0);
\coordinate (f) at ($(o)+(60:1.3)$);
\coordinate (b) at ($(o)+(180:2)$);

\draw[thick] (o) -- (60:2);
\draw[thick] (o) -- (-60:2);
\draw[thick] (o) -- (180:2);
\draw[thick, dashed] (f) -- +(-30:1.5);

\filldraw[thick, pattern = north east lines, rotate=45, pattern color = gray] (b) rectangle +(-2,2);

\node[cyan] () at ($(o)+(60:2.5)$){$1$};
\node[cyan] () at ($(f)+(-30:1.7)$){$z$};
\node[cyan] () at ($(o)+(-60:2.5)$){$0$};

\node[red] () at ($(o)+(60:1.5)+(150:0.3)$){$p_1$};
\node[red, scale=0.8] () at ($(o)+(60:0.5)+(-30:0.7)$){$p_1+\frac{\epsilon' b}{2}$};
\node[red] () at ($(f)+(30:1.2)+(-60:0.7)$){$p_{(1,2)}$};
\node[red] () at ($(o)+(-60:1)+(-150:0.3)$){$p_0$};
\node[red] () at ($(o)+(180:1)+(-90:0.4)$){$p_\sigma$};

\end{tikzpicture}
    }},\qquad \qquad \epsilon,\epsilon'=\pm.
    \eeq
    with the \textit{same} fusion matrix $\mathsf F$. The box
    \raisebox{-8pt}{\scalebox{0.3}{\begin{tikzpicture}\filldraw[thick, pattern = north east lines, rotate=45, pattern color = gray] (0,0) rectangle +(-2,2); \end{tikzpicture}}}
     denotes \textit{any} fixed admissible tree with momenta assigned to edges.  This exact fusion relation for multipoint conformal blocks with degenerate fields will play a crucial role below. Let us also record the normalizations
    \begin{subequations}\label{OPEs}
    \begin{align}\label{zeroopeA}
   \vcenter{\hbox{ 
   		\begin{tikzpicture}[scale=0.7]

\coordinate (o) at (0,0);
\coordinate (f) at ($(o)+(-60:1.3)$);
\coordinate (b) at ($(o)+(180:2)$);

\draw[thick] (o) -- (60:2);
\draw[thick] (o) -- (-60:2);
\draw[thick] (o) -- (180:2);
\draw[thick, dashed] (f) -- +(30:1.5);

\filldraw[thick, pattern = north east lines, pattern color = gray, rotate=45] (b) rectangle +(-2,2);

\node[cyan] () at ($(o)+(60:2.5)$){$1$};
\node[cyan] () at ($(f)+(30:1.7)$){$z$};
\node[cyan] () at ($(o)+(-60:2.5)$){$0$};

\node[red] () at ($(o)+(60:1)+(150:0.3)$){$p_1$};
\node[red, scale=0.8] () at ($(o)+(-60:0.5)+(30:0.7)$){$p_0+\frac{\epsilon b}{2}$};
\node[red] () at ($(f)+(30:1.2)+(-60:0.7)$){$p_{(1,2)}$};
\node[red] () at ($(o)+(-60:1.8)+(-150:0.3)$){$p_0$};
\node[red] () at ($(o)+(180:1)+(-90:0.4)$){$p_\sigma$};

\end{tikzpicture}
   	}} = &\,\quad\;\;
   z^{\frac{1+b^2}{2}-\epsilon bp_0}\quad\Biggl[
   \vcenter{\hbox{ \begin{tikzpicture}[scale=0.7]

\coordinate (o) at (0,0);
\coordinate (b) at ($(o)+(180:2)$);

\draw[thick] (o) -- (60:2);
\draw[thick] (o) -- (-60:2);
\draw[thick] (o) -- (180:2);

\filldraw[thick, pattern = north east lines, rotate=45, pattern color = gray] (b) rectangle +(-2,2);

\node[cyan] () at ($(o)+(60:2.5)$){$1$};
\node[cyan] () at ($(o)+(-60:2.5)$){$0$};

\node[red] () at ($(o)+(60:1)+(-30:0.3)$){$\quad p_1$};
\node[red, scale=1] () at ($(o)+(-60:1.2)+(30:1)$){$p_0+\frac{\epsilon b}{2}$};
\node[red] () at ($(o)+(180:1)+(-90:0.4)$){$p_\sigma$};

\end{tikzpicture}
   	}}\;+O\lb z\rb\,\Biggr]\qquad &\text{as }z\to0^+,\quad\\
  \;\; \vcenter{\hbox{ \begin{tikzpicture}[scale=0.7]

\coordinate (o) at (0,0);
\coordinate (f) at ($(o)+(60:1.3)$);
\coordinate (b) at ($(o)+(180:2)$);

\draw[thick] (o) -- (60:2);
\draw[thick] (o) -- (-60:2);
\draw[thick] (o) -- (180:2);
\draw[thick, dashed] (f) -- +(-30:1.5);

\filldraw[thick, pattern = north east lines, rotate=45, pattern color = gray] (b) rectangle +(-2,2);

\node[cyan] () at ($(o)+(60:2.5)$){$1$};
\node[cyan] () at ($(f)+(-30:1.7)$){$z$};
\node[cyan] () at ($(o)+(-60:2.5)$){$0$};

\node[red] () at ($(o)+(60:1.5)+(150:0.3)$){$p_1$};
\node[red, scale=0.8] () at ($(o)+(60:0.5)+(-30:0.7)$){$p_1+\frac{\epsilon' b}{2}$};
\node[red] () at ($(f)+(30:1.2)+(-60:0.7)$){$p_{(1,2)}$};
\node[red] () at ($(o)+(-60:1)+(-150:0.3)$){$p_0$};
\node[red] () at ($(o)+(180:1)+(-90:0.4)$){$p_\sigma$};

\end{tikzpicture}
  	}} =&\;
   \lb 1-z\rb^{\frac{1+b^2}{2}-\epsilon' b p_1}\Biggl[\vcenter{\hbox{
   		\begin{tikzpicture}[scale=0.7]

\coordinate (o) at (0,0);
\coordinate (b) at ($(o)+(180:2)$);

\draw[thick] (o) -- (60:2);
\draw[thick] (o) -- (-60:2);
\draw[thick] (o) -- (180:2);

\filldraw[thick, pattern = north east lines, rotate=45, pattern color = gray] (b) rectangle +(-2,2);

\node[cyan] () at ($(o)+(60:2.5)$){$1$};
\node[cyan] () at ($(o)+(-60:2.5)$){$0$};

\node[red] () at ($(o)+(60:1)+(-30:0.3)$){$\quad\qquad p_1+\frac{\epsilon' b}{2}$};
\node[red, scale=1] () at ($(o)+(-60:1.2)+(30:0.5)$){$p_0$};
\node[red] () at ($(o)+(180:1)+(-90:0.4)$){$p_\sigma$};

\end{tikzpicture} 
   	}}\;+O\lb z-1\rb\,\Biggr] &\text{as }z\to 1^-.\quad
    \end{align}
    \end{subequations}
    which are straightforward to obtain from the leading order of the OPE of the degenerate field with the relevant primary. In the above, it is implicitly assumed that $z\in \lb 0,1\rb$. Fractional powers for generic complex values are defined by analytic continuation.

     \subsection{Quasiclassical limit}
     Consider the limit where all external and internal momenta and the Virasoro central charge are sent to infinity according to
     \beq
    b\to 0, \qquad p_k\to 0,\qquad  bp_k\to\theta_k\quad \forall k.
     \eeq
     A famous conjecture of Zamolodchikov \cite{Zamo86} states that the corresponding behaviour of conformal blocks is given by
     \beq\label{zamoconj0}
     \ln \mathcal F\lb \left\{b^{-1}\theta_k\right\};\mathbf{t}\rb=b^{-2}\mathcal W\lb\left\{\theta_k \right\}; \mathbf{t}\rb+O\lb 1\rb\qquad\text{as }b\to0.
     \eeq
     This is a statement about the existence of the $b\to 0$ limit of the coefficients of the multivariate formal series $b^2\ln \mathcal F\lb \left\{b^{-1}\theta_k\right\};\mathbf{t}\rb$. The corresponding multivariate formal series $\mathcal W\lb\left\{\theta_k \right\}; \mathbf{t}\rb$ is called \textit{quasiclassical conformal block}. Global analytic structure of $\mathcal W\lb\left\{\theta_k \right\}; \mathbf{t}\rb$ lacks even a conjectural description already in the simplest 4-point case.
   
     The quasiclassical conformal blocks are conjecturally related to accessory parameters of ODEs of Heun type \cite{Zamo86,T10,LLNZ}. The link may be established by considering an additional degenerate field $\Phi_{(1,2)}\lb z\rb$. In the quasiclassical limit, its conformal dimension remains finite: $\Delta_{(1,2)}=-\frac12$ as $b\to 0$. Conformal block with such an extra insertion is expected to behave in the quasiclassical limit as
     \beq\label{zamoconj}
     \vcenter{\hbox{ \begin{tikzpicture}[scale=0.7]

\coordinate (o) at (0,0);
\coordinate (f) at ($(o)+(-60:1.3)$);
\coordinate (b) at ($(o)+(180:2)$);

\draw[thick] (o) -- (60:2);
\draw[thick] (o) -- (-60:2);
\draw[thick] (o) -- (180:2);
\draw[thick, dashed] (f) -- +(30:1.5);

\filldraw[thick, pattern = north east lines, pattern color = gray, rotate=45] (b) rectangle +(-2,2);

\node[cyan] () at ($(o)+(60:2.5)$){$1$};
\node[cyan] () at ($(f)+(30:1.7)$){$z$};
\node[cyan] () at ($(o)+(-60:2.5)$){$0$};

\node[red] () at ($(o)+(60:1)+(150:0.3)$){$p_1$};
\node[red, scale=0.8] () at ($(o)+(-60:0.5)+(30:0.7)$){$p_0+\frac{\epsilon b}{2}$};
\node[red] () at ($(f)+(30:1.2)+(-60:0.7)$){$p_{(1,2)}$};
\node[red] () at ($(o)+(-60:1.8)+(-150:0.3)$){$p_0$};
\node[red] () at ($(o)+(180:1)+(-90:0.4)$){$p_\sigma$};

\end{tikzpicture}
     	}} = \Psi_{\epsilon}\lb z;\mathbf{t}\rb \exp\left\{{b^{-2}\mathcal W\lb \mathbf{t}\rb}\right\}\Bigl[\,1+o\lb 1\rb\,\Bigr]\qquad \text{as }b\to 0.
     \eeq
     Substituting this asymptotics into BPZ equation, the latter transforms into an ODE,
     \beq\label{genheun}
     \left[\frac{d^2}{dz^2}+\sum_{k=0}^{n-2}\frac{\delta_k}{\lb z-t_k\rb^2}+\frac{\delta_{n-1}-\sum_{k=0}^{n-2}\delta_k}{z\lb z-1\rb}+\sum_{k=2}^{n-2}\frac{\lb t_k-1\rb \mathcal E_k}{z\lb z-1\rb\lb z-t_k\rb}\right]
     \Psi_\pm\lb z\rb=0,
     \eeq
     with 
     \begin{subequations}
     	\begin{align}
     \delta_k=\tfrac14-\theta_k^2,\qquad\qquad  &k=0,\ldots,n-1,\\ 
     \label{accmon}
     \mathcal E_k=t_k\,\frac{\partial \mathcal W}{\partial t_k},\qquad\qquad  & k=2,\ldots,n-2.
     \end{align}
     \end{subequations}
     Equation \eqref{genheun} is the normal form of the most general linear 2nd order ODE with $n$ Fuchsian singularities. Rescaled external momenta are directly related to local monodromy exponents $\frac12\pm \theta_k$ at the singular points; ${n-3}$ rescaled internal momenta such as $\sigma=bp_\sigma$ encode exponents of composite monodromy and parameterize accessory parameters $\mathcal E_{2},\ldots,\mathcal E_{n-2}$. 
     
     \subsection{Trieste connection formula}
     The structure of the operator product expansions encoded in the conformal block diagrams implies that the amplitudes $\Psi_{\pm}\lb z\rb$ have $z\to 0$ expansions of the form
     \beq
     \Psi_{\pm}\lb z\rb=\mathcal N_{\pm}z^{\frac12\mp \theta_0}\left[1+\sum_{k=1}^\infty \psi_{\pm,k}^{[0]}z^k\right].
     \eeq
     Therefore, these amplitudes give a basis of Frobenius solutions of  \eqref{genheun}. The normalization coefficients $\mathcal N_{\pm}$ can be determined from the leading OPE term in \eqref{zeroopeA}:
     \beq\label{normpref}
     \mathcal N_{\epsilon}=\lim_{b\to 0}\vcenter{\hbox{
     		\begin{tikzpicture}[scale=0.7]

\coordinate (o) at (0,0);
\coordinate (b) at ($(o)+(180:2)$);

\draw[thick] (o) -- (60:2);
\draw[thick] (o) -- (-60:2);
\draw[thick] (o) -- (180:2);

\filldraw[thick, pattern = north east lines, rotate=45, pattern color = gray] (b) rectangle +(-2,2);

\node[cyan] () at ($(o)+(60:2.5)$){$1$};
\node[cyan] () at ($(o)+(-60:2.5)$){$0$};

\node[red] () at ($(o)+(60:1)+(-30:0.3)$){$\quad p_1$};
\node[red, scale=1] () at ($(o)+(-60:1.2)+(30:1)$){$p_0+\frac{\epsilon b}{2}$};
\node[red] () at ($(o)+(180:1)+(-90:0.4)$){$p_\sigma$};

\end{tikzpicture}
     	}} \;\exp\left\{{b^{-2}\mathcal W\lb \mathbf{t}\rb}\right\} = \mathcal N\;\exp\left\{\frac{\epsilon}2 \frac{\partial \mathcal W}{\partial \theta_0}\right\}, \qquad \epsilon=\pm.
     \eeq
     Here $\mathcal N$ denotes the limit 
     \beq\label{defmathcalN}
     \mathcal N=\lim_{b\to 0}\vcenter{\hbox{ 	\begin{tikzpicture}[scale=0.7]

\coordinate (o) at (0,0);
\coordinate (b) at ($(o)+(180:2)$);

\draw[thick] (o) -- (60:2);
\draw[thick] (o) -- (-60:2);
\draw[thick] (o) -- (180:2);

\filldraw[thick, pattern = north east lines, rotate=45, pattern color = gray] (b) rectangle +(-2,2);

\node[cyan] () at ($(o)+(60:2.5)$){$1$};
\node[cyan] () at ($(o)+(-60:2.5)$){$0$};

\node[red] () at ($(o)+(60:1)+(-30:0.3)$){$\quad p_1$};
\node[red, scale=1] () at ($(o)+(-60:1.2)+(30:0.5)$){$p_0$};
\node[red] () at ($(o)+(180:1)+(-90:0.4)$){$p_\sigma$};

\end{tikzpicture}
     	}} \;\exp\left\{{b^{-2}\mathcal W\lb \mathbf{t}\rb}\right\},
     \eeq
    which is related to the $O\lb 1\rb$ correction in \eqref{zamoconj0}. We will not attempt to derive a more explicit expression of $\mathcal N$ since this prefactor is inessential for our purposes. 
     
     Indeed, we could repeat the same analysis for the quasiclassical limit of conformal blocks appearing in the right hand side of \eqref{exactfusion}. The corresponding amplitudes give a basis of Frobenius solutions of \eqref{genheun} at the singular point $z=1$. The analogs of the normalization prefactors \eqref{normpref} for these solutions are $\mathcal N \,\exp\left\{\frac{\epsilon'}2\frac{\partial \mathcal W}{\partial\theta_1}\right\}$ with the same $\mathcal N$ defined by \eqref{defmathcalN}. Given the exact fusion relation \eqref{exactfusion}, the only missing ingredient needed to derive the connection formula between the two bases is the quasiclassical limit of the fusion matrix \eqref{fusionmatr}. It is obviously given by 
    \beq\label{Fcl}
    \mathsf F_{\text{cl}}\lb \theta_0,\theta_1,\theta_\infty\rb=\frac{\Gamma\lb 1-2  \theta_0\rb\Gamma\lb2\theta_1\rb}{\Gamma\lb\tfrac12+\theta_1-\theta_0+\theta_\infty\rb \Gamma\lb\tfrac12+\theta_1-\theta_0-\theta_\infty\rb}.
    \eeq
    
    This yields the following statement. Let us denote by $\psi^{[0]}_{\pm}\lb z\rb$, $\psi^{[1]}_{\pm}\lb z\rb$ two pairs of \textit{normalized} Frobenius solutions of the generalized Heun equation \eqref{genheun} at the points $z=0$ and $z=1$:
    \beq\label{frobases}
    \psi^{[0]}_{\pm}\lb z\rb=z^{\frac12\mp\theta_0}\left[1+\sum_{k=1}^\infty \psi^{[0]}_{\pm,k} z^k\right],\qquad 
    \psi^{[1]}_{\pm}\lb z\rb=\lb 1-z\rb^{\frac12\mp\theta_1}\left[1+\sum_{k=1}^\infty \psi^{[1]}_{\pm,k} \lb z-1\rb^k\right].
    \eeq 
    The connection between the two bases is given by 
    \begin{subequations}
    	\label{triestef}
    \begin{align}	
    &\label{triestefA}\psi^{[0]}_{\epsilon}\lb z\rb=\sum_{\epsilon'}\;\mathsf{C}\lb \epsilon\theta_0,\epsilon'\theta_1,\sigma\rb
    \psi^{[1]}_{\epsilon'}\lb z\rb,\qquad \epsilon,\epsilon'=\pm,\\
    &\boxed{\mathsf{C}\lb\theta_0,\theta_1,\sigma\rb=\mathsf{F}_{\text{cl}}\lb \theta_0,\theta_1,\sigma\rb \exp\frac{1}2\lb\frac{\partial \mathcal W}{\partial\theta_1}-\frac{\partial \mathcal W}{\partial\theta_0}\rb}
    \end{align}
    \end{subequations}
    An equivalent result was first obtained in the 4-point case (the usual Heun equation) in \cite{BILT22}, cf. eqs. (4.1.16)--(4.1.17) therein. We will refer to it as the \textit{Trieste formula}.
    
    The dressing of the quasiclassical fusion matrix $\mathsf F_{\text{cl}}$ by conformal block derivatives  $\frac{\partial \mathcal W}{\partial\theta_{0}}$, $\frac{\partial \mathcal W}{\partial\theta_{1}}$ is a consequence of the shifts of the external momenta by $\pm\frac b2$ in the leading OPE terms in \eqref{OPEs}. Connection matrices between Frobenius solutions at $z=0$ and other singular points $t_2,\ldots, t_{n-1}$ can be computed in an analogous way using along with fusion the elementary braiding transformations. In general, the transformed conformal blocks also involve shifts of the internal momenta by half-integer multiples of $b$, which in the quasiclassical limit produce derivatives with respect to rescaled internal momenta. 
    
    The monodromy of solutions of \eqref{genheun} can be written in terms of the connection matrices. A class of monodromy invariants has a particularly simple expression which does not involve  conformal block functions explicitly. For example, the spectrum of the monodromy matrix $M_0M_1$ around two singular points $z=0,1$ is given by $\operatorname{Sp}M_0M_1=\left\{-e^{\pm2\pi i \sigma}\right\}$. In a similar fashion, the other rescaled internal momenta are related to exponents of composite monodromy along the cycles encoded in the conformal block diagram interpreted as a pants decomposition of $\mathbb C\mathbb P^1\backslash\mathbf{t}$.
       
     Under minimal adjustments, the above argument can also be carried out for irregular conformal blocks leading to  connection formulas for a number of confluent Heun equations \cite{BILT21,BILT22}. We also note a recent work \cite{JN} where results related to \eqref{triestef} were derived on the gauge theory side of the AGT correspondence (cf e.g. formulas of Subsection~6.1 therein).
       
    \subsection{Practical implementation for Heun equation}
     An important subtlety with the application of Trieste formula is that the connection coefficients are expressed in terms of monodromy invariants (such as $\sigma$) instead of parameters $\left\{\mathcal E_k\right\}$ of the generalized Heun equation. The connection between the two is given by the equation \eqref{accmon} that has to be solved perturbatively.
     
      For reader's convenience, let us outline the procedure in the case of the usual (4-point) Heun equation written in the normal form
      \beq\label{usheun}
     \left[\frac{d^2}{dz^2}+\frac{\delta_0}{z^2}+\frac{\delta_1}{\lb z-1\rb^2}+\frac{\delta_t}{\lb z-t\rb^2}+\frac{\delta_\infty-\delta_0-\delta_1-\delta_t}{z\lb z-1\rb}+\frac{\lb t-1\rb \mathcal E}{z\lb z-1\rb\lb z-t\rb}\right]
     \psi\lb z\rb=0,
     \eeq
     where $\delta_k=\frac14-\theta_k^2$ for $k=0,1,t,\infty$ and $1<|t|<\infty$.  The perturbative expansion of the quasiclassical conformal block $\mathcal W\lb t\rb$ in $1/t$  has the form
     \beq\label{clcbexp}
     \mathcal W\lb t\rb= \lb \delta_\infty-\delta_\sigma-\delta_t\rb\,\ln t+\sum_{k=1}^\infty\mathcal W_k t^{-k}.
     \eeq
     Here $\delta_\sigma=\frac14-\sigma^2$ and the first coefficients are given by
     \begin{subequations}
     	\label{Wcoefs}
     \begin{align}
     &\mathcal W_1=\frac{\lb \delta_\sigma-\delta_0+\delta_1\rb \lb \delta_\sigma-\delta_\infty+\delta_t\rb}{2\delta_\sigma},\\
     \label{2ndcoef}
    &\begin{aligned}\mathcal W_2=& \frac{\lb \delta_\sigma-\delta_0+\delta_1\rb^2 \lb \delta_\sigma-\delta_\infty+\delta_t\rb^2}{8\delta_\sigma^2}\left(\frac{1}{\delta_\sigma-\delta_0+\delta_1}+\frac{1}{\delta_\sigma-\delta_\infty+\delta_t}-\frac{1}{2\delta_\sigma}\right)+
    \\
     &\quad +\frac{\lb\delta_\sigma^2+2\delta_\sigma\lb\delta_0+\delta_1\rb-
     	3\lb\delta_0-\delta_1\rb^2\rb \lb\delta_\sigma^2+2\delta_\sigma\lb\delta_\infty+\delta_t\rb-3\lb\delta_\infty-\delta_t\rb^2\rb}{16\delta_\sigma^2\lb 4\delta_\sigma+3\rb}.
     \end{aligned}
     \end{align}
     \end{subequations}
     The calculation of subsequent coefficients can be carried out by using any suitable method (e.g. Zamolodchikov recursion \cite{Zamo84} or Nekrasov functions \cite{Nekrasov}) to generate conformal block expansion  and computing the quasiclassical limit thereof. 
     
     Parameterize the accessory parameter as
     \beq
     \mathcal E=\delta_\infty-\delta_{\omega}-\delta_t=-\tfrac14-\theta_\infty^2+\omega^2+\theta_t^2.
     \eeq
     The meaning of the new parameter $\omega$ introduced instead of $\mathcal E$ is the limiting value of the composite monodromy exponent $\sigma$ as $t\to\infty$.
     Indeed, Zamolodchikov relation $\mathcal E=t\frac{\partial\mathcal W}{\partial t}$ leads to the perturbative series
     \beq
     \sigma\lb t\rb=\omega+\sum_{k=1}^\infty\sigma_kt^{-k}.
     \eeq
     For any $k$, the coefficient $\sigma_k$ is determined by $\mathcal W_1,\ldots,\mathcal W_k$ and is given by a rational function of $\boldsymbol\theta=\left\{\theta_0,\theta_1,\theta_t,\theta_\infty\right\}$ and $\omega$, e.g.
      \beq
     \sigma_1=\frac{\lb\frac14-\omega^2+\theta_0^2-\theta_1^2\rb
     	\lb\frac14-\omega^2+\theta_\infty^2-\theta_t^2\rb}{4\omega\lb\frac14 -\omega^2\rb}.
     \eeq
     
     Expanding the first factor in the Trieste formula   \eqref{triestef}, one can write 
     \beq\label{exp00}
     \ln\frac{\mathsf{F}_{\text{cl}}\lb \theta_0,\theta_1,\sigma\rb}{\mathsf{F}_{\text{cl}}\lb \theta_0,\theta_1,\omega\rb}= \sum_{k=0}^\infty\left[\lb -1\rb^{k}\psi^{(k)}\lb\tfrac12+\theta_1-\theta_0-\omega\rb-\psi^{(k)}\lb\tfrac12+\theta_1-\theta_0+\omega\rb\right]\frac{\lb \sigma-\omega\rb^{k+1}}{\lb k+1\rb!}.
     \eeq
     It then becomes clear that the coefficients $\mathsf f_k$ of the perturbative expansion 
     \beq
      \ln\mathsf{C}\lb \theta_0,\theta_1,\sigma\rb=\ln\mathsf{F}_{\text{cl}}\lb \theta_0,\theta_1,\omega\rb+\sum_{k=1}^{\infty}\mathsf f_k t^{-k}
     \eeq
     are given by linear combinations of polygamma functions $\psi^{(n)}\lb\tfrac12+\theta_1-\theta_0\pm\omega\rb$ ($n=0,\ldots,k-1$) with coefficients rational in $\boldsymbol\theta$ and $\omega$. In particular,
     \beq\label{coeff1}
     \begin{aligned}
     \mathsf f_1=&\,-\frac{\lb\frac14-\omega^2+\theta_0^2-\theta_1^2\rb
     	\lb\frac14-\omega^2+\theta_\infty^2-\theta_t^2\rb}{4\omega\lb\frac14 -\omega^2\rb}\Bigl[
     \psi\lb\tfrac12+\theta_1-\theta_0+\omega\rb-\psi\lb\tfrac12+\theta_1-\theta_0-\omega\rb\Bigr]
     \\
     &\,-\frac{\lb\theta_0+\theta_1\rb\lb\frac14-\omega^2+\theta_\infty^2-\theta_t^2\rb}{2\lb\frac14-\omega^2\rb}.
     \end{aligned}
     \eeq
     The expressions for $\sigma_2$ and $\mathsf{f}_2$ are omitted for the sake of brevity, yet 
     they are completely straightforward to compute from \eqref{clcbexp}-\eqref{Wcoefs} and \eqref{exp00}.

     \section{Schäfke-Schmidt connection formula}\label{sectionSS}
     Consider a more general linear ODE 
     \beq\label{Genheun}
     \left[\frac{d^2}{dz^2}+\frac{\frac14-\theta_0^2}{z^2}+\frac{\frac14-\theta_1^2}{\lb z-1\rb^2}+\frac{U\lb z\rb}{z\lb z-1\rb}\right]\psi\lb z\rb=0,
     \eeq
     where $U\lb z\rb$ is analytic inside the disk $|z|<R$ with $R>1$. The generalized Heun equation \eqref{genheun} is clearly a special case of \eqref{Genheun}. Throughout the rest of the paper, it is assumed that $\theta_0,\theta_1\notin\mathbb Z/2$, in which case there exist unique bases of normalized Frobenius solutions $\psi_{\pm}^{[0]}\lb z\rb$, $\psi_{\pm}^{[1]}\lb z\rb$ whose expansions near the Fuchsian singularities $z=0,1$ have the form \eqref{frobases} .
     
     The Wronskian $W\lb\psi_a,\psi_b\rb=\psi_a\psi_b'-\psi'_a\psi_b$ of any pair of solutions of \eqref{Genheun} is independent of $z$. It is straightforward to check that
     \beq\label{w0011}
     W\lb\psi_+^{[0]},\psi_-^{[0]}\rb=2\theta_0,\qquad 
     W\lb\psi_+^{[1]},\psi_-^{[1]}\rb=-2\theta_1.
     \eeq
     The connection formula
     \beq\label{conngen}
     \psi^{[0]}_{\epsilon}\lb z\rb=\sum_{\epsilon'}\;\mathsf{C}_{\epsilon\epsilon'}
     \psi^{[1]}_{\epsilon'}\lb z\rb,\qquad\qquad \epsilon,\epsilon'=\pm
     \eeq
     then implies that
     \beq\label{Cwr}
     \mathsf C_{\epsilon\epsilon'}=\mathsf C\lb \epsilon\theta_0,\epsilon'\theta_1\rb,\qquad
     \mathsf C\lb \theta_0,\theta_1\rb=-\tfrac1{2\theta_1}W\lb\psi_+^{[0]},\psi_-^{[1]}\rb.
     \eeq
     The connection coefficients are thus expressed in terms of a single function $\mathsf C\lb \theta_0,\theta_1\rb$, just as in \eqref{triestefA} above. This is a consequence of the invariance of \eqref{Genheun} 
     with respect to the sign flips $\theta_0\mapsto-\theta_0$, $\theta_1\mapsto-\theta_1$ of local monodromy exponents. 
     
     From \eqref{w0011} and \eqref{conngen} it follows that 
     $ \operatorname{det} \mathsf C=-\ds\frac{\theta_0}{\theta_1}$.
     Expressing the composite monodromy around $0$ and $1$ in terms of $\mathsf C_{\epsilon\epsilon'}$, we can also write
     \beq
     \operatorname{Tr}\lb\begin{array}{rr}
     	\mathsf C_{++} & \mathsf C_{+-} \\ \mathsf C_{-+} & \mathsf C_{--}
     	\end{array}\rb \lb\begin{array}{cc}
     	e^{2\pi i\theta_1} & 0 \\ 0 & e^{-2\pi i\theta_1}
     \end{array}\rb \lb\begin{array}{rr}
     	\mathsf C_{--} & -\mathsf C_{+-} \\ -\mathsf C_{-+} & \mathsf C_{++}
     \end{array}\rb 
 \lb\begin{array}{cc}
 	e^{2\pi i\theta_0} & 0 \\ 0 & e^{-2\pi i\theta_0}
 \end{array}\rb=-2\cos2\pi\sigma\cdot\operatorname{det} \mathsf C.
     \eeq 
     This yields the relations 
     \beq
     \mathsf C_{+\pm}\mathsf C_{\mp-}=-\frac{\theta_0}{\theta_1}\frac{\cos\pi\lb\theta_1\mp\theta_0+\sigma\rb \cos\pi\lb\theta_1\mp\theta_0-\sigma\rb}{\sin2\pi\theta_0\sin2\pi\theta_1}.
     \eeq
     It is a simple exercise to check that these two identities are satisfied by the conjectural formula \eqref{triestef} of the previous section.
     
     Equation \eqref{Cwr} is already sufficient for numerical evaluation of the connection coefficients $\mathsf C_{\epsilon\epsilon'}$. Indeed, using \eqref{Genheun} one may generate the expansions \eqref{frobases} truncated at sufficiently large order, and then compute the Wronskians at an intermediate point $\tilde z\in(0,1)$. Our analytic perturbative calculation for Heun equations  will be based on a result of Schäfke and Schmidt \cite{SS} which relates $ \mathsf  C\lb\theta_0,\theta_1\rb$ to the asymptotics of expansion coefficients of suitably modified Frobenius solutions.
     
     We start by recalling a statement which originates from a classical work of Darboux \cite[p. 10]{Darboux}.
     \begin{prop}
     	Let $u\lb z\rb$ be a function having exactly one branch point $z=1$ inside a circle $|z|=R>1$. If $u\lb z\rb$ can be represented as
     	\beq\label{uvw}
     	u\lb z\rb=v\lb z\rb+\lb 1-z\rb^{-\theta}w\lb z\rb,
     	\eeq 
     with $v\lb z\rb$, $w\lb z\rb$ analytic in a neighborhood of $z=1$, then the coefficients of the Taylor expansion $u\lb z\rb=\sum_{k=0}^\infty u_kz^k$ at $z=0$ have the asymptotics
     \beq\label{ukas}
     u_k=w\lb 1\rb\frac{k^{\theta-1}}{\Gamma\lb \theta\rb}\left[1+o\lb 1\rb\right]\qquad \text{as }k\to\infty.
     \eeq	
     \end{prop}
     \pf Consider the contour $\mathcal C=\mathcal C_r\cup\mathcal C_+\cup\mathcal C_-\cup\mathcal C_{R'}$ represented in Fig.~1a, where $R'<R$ and $r$ is chosen sufficiently small so that $v\lb z\rb$, $w\lb z\rb$ are analytic inside an open disk containing  $\mathcal C_r$. 
     \begin{figure}[h!]
     	\begin{center}
    \begin{tikzpicture}[scale=1]

\coordinate (o) at (1,0);
\coordinate (o2) at ($(o)+(0:1.5)$);

\filldraw[black] (o) circle (1pt);
\filldraw[black] (o2) circle (1pt);

\node[cyan] () at ($(o)+(0,-0.3)$){$0$};
\node[red] () at ($(o)+(240:1.5)$){$R'$};
\node[red] () at ($(o)+(160:2.6)$){$\mathcal{C}_{R'}$};

\draw (o)+(0:-3) arc (180:5:3 and 3);
\draw (o)+(0:-3) arc (-180:-5:3 and 3);
\draw[-{Stealth[scale=2]}] (o) -- +(230:3);
\draw[-{To[scale=2]}] ($(o)+(90:3)$) -- +(0:-0.01);

\node[cyan] () at ($(o2)+(0,-0.3)$){$1$};
\node[red] () at ($(o2)+(210:0.5)$){$r$};
\node[red] () at ($(o2)+(240:1.3)$){$\mathcal{C}_{r}$};

\draw (o2)+(0:-1) arc (180:15:1 and 1);
\draw (o2)+(0:-1) arc (-180:-15:1 and 1);
\draw[-{Stealth[scale=1.5]}] (o2) -- +(230:1);
\draw[-{To[scale=2]}] ($(o2)+(90:1)$) -- +(0:0.01);

\draw[-{To[scale=1.2]}] ($(o2)+(15:1)$) -- +(0:0.3);
\draw ($(o2)+(15:1)+(0:0.25)$) -- +(0:0.28);
\draw[-{To[reversed,scale=1.2]}] ($(o2)+(-15:1)$) -- +(0:0.3);
\draw ($(o2)+(-15:1)+(0:0.25)$) -- +(0:0.28);

\node[scale=0.8, red] () at ($(o2)+(0:1.2)+(0, 0.5)$){$\mathcal{C}_+$};
\node[scale=0.8, red] () at ($(o2)+(0:1.2)+(0,-0.5)$){$\mathcal{C}_-$};

\node () at ($(o)+(0,-3.5)$){$(a)$};

\coordinate (o) at (7,0);
\coordinate (o2) at ($(o)+(0:1.5)$);

\filldraw[black] (o) circle (1pt);
\filldraw[black] (o2) circle (1pt);

\node[cyan] () at ($(o)+(-0.3,-0.3)$){$0$};
\node[cyan] () at ($(o2)+(0,-0.3)$){$1$};
\node[cyan] (inf) at ($(o2)+(4.5,0)$){$\infty$};
\node[red] () at ($(o2)+(240:1.3)$){$\mathcal{C}_r$};

\draw[decoration = {zigzag,segment length = 1mm, amplitude = 0.3mm},decorate, black!60!red] (o2) -- (inf);
\draw (o2)+(0:-1) arc (180:15:1 and 1);
\draw (o2)+(0:-1) arc (-180:-15:1 and 1);
\draw[-{To[scale=2]}] ($(o2)+(90:1)$) -- +(0:0.01);

\draw[-{To[scale=1.8]}] ($(o2)+(15:1)$) -- +(0:2);
\draw ($(o2)+(15:1)+(0:2)$) -- +(0:1.5);
\draw[-{To[reversed,scale=1.8]}] ($(o2)+(-15:1)$) -- +(0:2);
\draw ($(o2)+(-15:1)+(0:1.5)$) -- +(0:2);

\node[scale=0.8, red] () at ($(o2)+(0:1.2)+(1.5, 0.7)$){$\mathcal{C}_+$};
\node[scale=0.8, red] () at ($(o2)+(0:1.2)+(1.5,-0.7)$){$\mathcal{C}_-$};

\node () at ($(o)+(3,-3.5)$){$(b)$};

\end{tikzpicture} \\
     	Fig.~1: Integration contours.
     	\end{center}
     \end{figure}
     Clearly, $u_k=\ds\frac1{2\pi i}\ds\oint_{\mathcal C}z^{-k-1}u\lb z\rb dz$.  The integrals along $\mathcal C_{R'}$ and $\mathcal C_{\pm}$ are exponentially suppressed as $O\lb R'^{-k}\rb$ and $O\lb \lb 1+r\rb^{-k}\rb$ as $k\to\infty$. Substituting \eqref{uvw} into the remaining integral along $\mathcal C_r$, we further notice that the contribution of $v\lb z\rb$ vanishes. The contribution of $w\lb z\rb$ can be estimated as
     \beq
     \ds\frac{w\lb 1\rb}{2\pi i}\ds\oint_{\mathcal C_r}z^{-k-1}\lb1-z\rb^{-\theta}  dz\;\substack{k\to\infty\\ = \\ \;}\; 
     \ds\frac{w\lb 1\rb}{2\pi i}\ds\oint_{\mathcal C_r\cup\mathcal C_+'\cup\mathcal C_-'}z^{-k-1}\lb1-z\rb^{-\theta}  dz=
     \frac{\Gamma\lb k+\theta\rb}{\Gamma\lb k+1\rb\Gamma\lb\theta\rb}\,w\lb 1\rb,
     \eeq 
     where $\mathcal C_\pm'$ denote the horizontal segments depicted in Fig.~1b. The asymptotics \eqref{ukas} then follows from the Stirling's approximation for the gamma function.
     \epf
     \begin{cor}[Schäfke-Schmidt formula] \label{CorSS}
     	Connection matrix relating the normalized Frobenius bases of solutions of \eqref{Genheun} at $z=0,1$ is given by
     	\beq\label{SSfor}
     	\mathsf C\lb\theta_0,\theta_1\rb=\Gamma\lb2\theta_1\rb\lim_{k\to\infty}k^{1-2\theta_1}u_k,
     	\eeq
     	where $u_k$ denote the coefficients of the Taylor expansion at $0$ of the function
     	\beq\label{defuz}
     	u\lb z\rb:=z^{-\frac12+\theta_0}\lb 1-z\rb^{-\frac12-\theta_1}\psi_{+}^{[0]}\lb z\rb=1+\sum_{k=1}^\infty u_kz^k.
     	\eeq
     \end{cor}
     \pf The conditions of the previous proposition are verified with $u\lb z\rb$ defined by \eqref{defuz}, $\theta=2\theta_1$ and
     \begin{subequations}
     \begin{align}
     v\lb z\rb=&\,\mathsf C\lb\theta_0,-\theta_1\rb z^{-\frac12+\theta_0}
     \lb 1-z\rb^{-\frac12-\theta_1}\psi_{-}^{[1]}\lb z\rb, \\
     w\lb z\rb=&\,\;\;\mathsf C\lb\theta_0,\theta_1\rb\;\; z^{-\frac12+\theta_0}\lb 1-z\rb^{-\frac12+\theta_1}\psi_{+}^{[1]}\lb z\rb.
     \end{align} 
     \end{subequations}
     Indeed, $v\lb z\rb$ and $w\lb z\rb$ are holomorphic in a neighborhood of $z=1$ and $w\lb 1\rb=\mathsf C\lb\theta_0,\theta_1\rb$. The relation \eqref{uvw} comes from \eqref{conngen} with $\epsilon=+$. \epf

     Let us illustrate the formula \eqref{SSfor} with a toy example of the Gauss hypergeometric equation. The latter corresponds to the choice  $U\lb z\rb=\theta_0^2+\theta_1^2-\theta_\infty^2-\frac14$. The ODE \eqref{Genheun} for $\psi_{+}^{[0]}\lb z\rb$ then  transforms into the canonical form of the hypergeometric equation satisfied by $u\lb z\rb$:
     \beq\label{hypercan}
     \left[z\lb 1-z\rb\frac{d^2}{dz^2}+\bigl( 1-2\theta_0-2\lb 1-\theta_0+\theta_1\rb z\bigr)\,\frac{d}{dz}+\theta_\infty^2-\lb\tfrac12-\theta_0+\theta_1\rb^2\right]u\lb z\rb=0.
     \eeq
     This in turn yields a 2-term linear recurrence relation 
     \beq\label{2termHG}
     u_{k+1}=\frac{\lb\tfrac12-\theta_0+\theta_1+k\rb^2-\theta_\infty^2}{\lb k+1\rb\lb k+1-2\theta_0\rb}\,u_k.
     \eeq
     Given the initial condition $u_0=1$, its solution reads
     \beq
     u_k=\frac{\lb \tfrac12-\theta_0+\theta_1+\theta_\infty\rb_k \lb \tfrac12-\theta_0+\theta_1-\theta_\infty\rb_k}{k!\,\lb 1-2\theta_0\rb_k},
     \eeq
     where $\lb x\rb_k=\frac{\Gamma\lb x+k\rb}{\Gamma\lb x\rb}$ denotes the Pochhammer symbol. Substituting this expression into \eqref{SSfor} and computing the limit, one finds that
     $\mathsf C\lb\theta_0,\theta_1\rb=\mathsf F_{\text{cl}}\lb\theta_0,\theta_1,\theta_\infty\rb$,
     with $\mathsf F_{\text{cl}}$ given by \eqref{Fcl}. We thereby recover the standard hypergeometric connection formulas.

     \section{Application to Heun equations}\label{sectionHeun}
     \subsection{Reduced confluent Heun equation (RCHE)}
     We now proceed to the connection problem for Heun equations with at least two Fuchsian singularities: reduced confluent, confluent and usual Heun equation. To explain the idea with a minimal amount of fuss, it is convenient to start with the reduced confluent case. This corresponds to setting in \eqref{Genheun}
     \beq
     U\lb z\rb=\theta_0^2+\theta_1^2-\omega^2-\tfrac14-\lambda z,
     \eeq
     where $\omega$ plays the role of accessory parameter and $\lambda$ is a coupling constant. We will be mainly interested in the weak coupling regime $|\lambda|\ll 1$.
     
     The function $u\lb z\rb$ defined by \eqref{defuz} satisfies the canonical form of the RCHE:
      \beq\label{RCHEcan}
     \left[z\lb 1-z\rb\frac{d^2}{dz^2}+\bigl( 1-2\theta_0-2\lb 1-\theta_0+\theta_1\rb z\bigr)\,\frac{d}{dz}+\omega^2-\lb\tfrac12-\theta_0+\theta_1\rb^2+\lambda z\right]u\lb z\rb=0.
     \eeq
     \begin{lemma}\label{lemmaRCHE1}
     	The connection matrix relating Frobenius solutions of the RCHE at $z=0,1$ is given by
     	\beq
     	\mathsf C\lb\theta_0,\theta_1\rb=\frac{\Gamma\lb 1-2  \theta_0\rb\Gamma\lb2\theta_1\rb}{\Gamma\lb\tfrac12+\theta_1-\theta_0+\omega\rb \Gamma\lb\tfrac12+\theta_1-\theta_0-\omega\rb}\, a_\infty,
     	\eeq
     	where  $a_\infty=\lim\limits_{k\to\infty}a_k$ is the limit of the sequence defined by the 3-term recurrence relation
     	\beq\label{3termRCHE}
     	a_{k+1}-a_k=-\lambda\beta_k a_{k-1},
     	\eeq
     with $a_{-1}=0$, $a_0=1$ and $\beta_k$ defined by \eqref{betaRCHE}.
     \end{lemma}
     \pf Equation \eqref{RCHEcan} implies the following 3-term recurrence for the coefficients $u_k$: 
     \beq
    \lb k+1\rb\lb k+1-2\theta_0\rb u_{k+1}=
    \lb\lb\tfrac12-\theta_0+\theta_1+k\rb^2-\theta_\infty^2\rb\,u_k-\lambda u_{k-1},
     \eeq
     which in the case $\lambda=0$ reduces to \eqref{2termHG}. Introducing rescaled coefficients $a_k={u_k}/{u_k^{\lambda=0}}$, it is straightforward to show that they satisfy \eqref{3termRCHE}, \eqref{betaRCHE}. The statement of the lemma then follows from Corollary~\ref{CorSS} and the hypergeometric example in the end of the previous section.
     \epf
     
     In order to find the perturbative expansion of $a_\infty$ for small $\lambda$, we first
     rewrite it as a determinant of an infinite tridiagonal matrix.
     \begin{lemma}
     	The limiting value $a_\infty$ can be written as
     \beq\label{infdetRCHE}
     a_\infty=\operatorname{det}\lb\begin{array}{ccccc}1 & -1 &  & \\
     	-\lambda\beta_1 & 1 & -1 &  & \\
     	& -\lambda\beta_2 & 1 & -1 & \\
     	& & -\lambda\beta_3 & 1 & \cdot\\
     	& & & \cdot & \cdot 
     \end{array}\rb.
     \eeq	
     \end{lemma}
     \pf Interpret the recurrence relations \eqref{3termRCHE} with $k=0,\ldots,n-1$ in combination with the initial condition $a_0=1$ as a linear system for $a_0,\ldots,a_n$. Solving it for $a_n$ and sending $n\to\infty$ gives the representation \eqref{infdetRCHE}. \epf
     
     One could attempt to write a formal power series expansion
     \beq
     a_\infty=1-\lambda\sum_{k=1}^\infty \beta_k+\lambda^2\sum_{k=1}^{\infty}\sum_{k'=k+2}^\infty\beta_k\beta_{k'}+\ldots
     \eeq
     It is however more efficient to expand $\ln a_\infty$ as this eliminates nested sums. 
     \begin{lemma}
     We have
     \beq\label{lndet}
     \ln a_\infty=-\sum_{n=1}^\infty \frac{\operatorname{Tr}\,A^{2n}}{2n}\, \lambda^n,
     \eeq 
     where $A$ denotes the infinite matrix 
     \beq
     A=\lb\begin{array}{ccccc}0 & 1 &  & \\
     	\beta_1 & 0 & 1 &  & \\
     	& \beta_2 & 0 & 1 & \\
     	& & \beta_3 & 0 & \cdot\\
     	& & & \cdot & \cdot 
     \end{array}\rb.
     \eeq
     \end{lemma}
 
      We have, in particular,
      \begin{subequations}\label{expltraces}
     \begin{align}
     \label{expltracesa}\operatorname{Tr}{A^2}=&\sum_{k=1}^{\infty}2\beta_k,\qquad
     \operatorname{Tr}{A^4}=\sum_{k=1}^{\infty}\lb 4\beta_k\beta_{k+1}+2\beta_k^2\rb, \\
     \label{tracesb}
     \operatorname{Tr}{A^6}=&\sum_{k=1}^{\infty}\lb 6\beta_k\beta_{k+1}\beta_{k+2}+6\beta_k^2\beta_{k+1}+6\beta_k\beta_{k+1}^2+2\beta_k^3\rb,\quad \ldots
     \end{align}
     \end{subequations}
 
     The general expression for $\operatorname{Tr}{A^{2n}}$ in terms of $\beta_k$'s can be written in a combinatorial way. Consider a staircase walk $W$ on the $n\times n$ square grid going from $\lb 0,0\rb$ to $\lb n,n\rb$ as shown in the example in Fig.~2a. The vertical edges of $W$ intersect $\ell\leq n$ consecutive diagonals parallel to the main one and passing through the midpoints of the edges. Order these diagonals in the northwest direction and denote the corresponding intersection numbers by $\mu_1,\ldots,\mu_\ell$. Obviously, $\mu_1+\ldots+\mu_\ell=n$. In this way we assign to every walk its type $\mu=\lb\mu_1,\ldots,\mu_\ell\rb$ --- a composition (ordered partition) of $n$ into $\ell$ parts. It will be denoted $\mu\vdash n$ similarly to ordinary partitions. The total number of compositions of $n$ is known to be equal to $2^{n-1}$. We may then write
     \beq\label{combiN}
     \operatorname{Tr}A^{2n}=\sum\limits_{k=1}^{\infty}\sum\limits_{\mu\vdash n}^{2^{n-1}} N_{\mu}\cdot\beta_{k}^{\mu_1}\beta_{k+1}^{\mu_2}\ldots\beta_{k+\ell}^{\mu_\ell},
     \eeq
     where $N_\mu$ denotes the number of staircase walks of type $\mu$. For example, the term $6\beta_k\beta_{k+1}^2$ in \eqref{tracesb} corresponds to six possible walks of type $(1,2)$ represented in Fig.~2b. The total number of staircase walks is $\sum_{\mu\vdash n}^{2^{n-1}} N_{\mu}={2n \choose n}$.
     
     \begin{figure}[h!]\label{figwalks}
     	\begin{center}
     		\newcommand{\gridd}[2]{
\coordinate (o) at #1;
\foreach \i in {0,...,#2} {
    \draw[thick] ($(o)+(0,\i)$) -- ($(o)+(#2,\i)$);
    \draw[thick] ($(o)+(\i,0)$) -- ($(o)+(\i,#2)$);}}
    
\newcommand{\pathh}[2]{
\coordinate (o1) at #1;
\foreach \i in #2{
    \coordinate (o2) at ($(o1)+({\i==0?1:0},\i)$);
    \draw[very thick, red] (o1) -- (o2);
    \coordinate (o1) at (o2);}}
      
\begin{tikzpicture}[scale=0.7]
\foreach \i in {0,4,8}{
    \gridd{(\i,0)}{3};
    \gridd{(\i,4)}{3};}

\pathh{(0,0)}{{0,1,0,0,2}}
\pathh{(4,0)}{{0,0,2,0,1}}
\pathh{(8,0)}{{1,0,0,2,0}}
\pathh{(0,4)}{{2,0,1,0,0}}
\pathh{(4,4)}{{1,0,1,0,0,1}}
\pathh{(8,4)}{{0,2,0,1,0}}

\node () at ($(5.5,-1.5)$){$(b)$};

\coordinate (b) at (-10, 0.5);

\gridd{(b)}{6}
\pathh{(b)}{{2,0,0,0,1,0,1,0,0,2}}
\foreach \x in {1,...,4}{
    \draw[thick, densely dashed, black!60!green] ($(b)+(0.5,-1)+(-\x/2+.5,\x/2-.5)$) -- ($(b)+(7,5.5)+(-\x/2+.5,\x/2-.5)$);
    \node[black!60!green] () at ($(b)+(7.2,5.7)+(-\x/2+.5,\x/2-.5)$){\x};}
\node[cyan] () at ($(b)+(-1,-0.6)$){$(0,0)$};
\node[cyan] () at ($(b)+( 7.5, 6.8)$){$(6,6)$};
\filldraw[blue] ($(b)+(0,0)$) circle (1pt);
\filldraw[blue] ($(b)+(6,6)$) circle (1pt);

\node () at ($(b)+(3,-2)$){$(a)$};
\end{tikzpicture}\\
     		Fig.~2: (a) A staircase walk of type $\lb 1,3,1,1\rb$; (b) Six possible walks of type $\lb 1,2\rb$.
     	\end{center}
           \end{figure}
     
     We now finish the proof of Theorem~\ref{thmA} from the Introduction and simultaneously obtain an explicit expression for the coefficients $N_\mu$.
     \begin{lemma}\label{contfrRCHElem}
     	The quantity $a_\infty$ admits the following infinite fraction representation:
     	\beq\label{contfrRCHE}
     	\ln a_\infty=\sum_{k=1}^\infty\ln\lb 1-\frac{\lambda\beta_k}{1-\frac{\lambda\beta_{k+1}}{1-\ldots}}\rb.
     	\eeq
     \end{lemma}
    \pf Consider a sequence of infinite determinants
     \beq
     D_k=\operatorname{det}\lb\begin{array}{cccc}1 & -1 &  & \\
     	-\lambda\beta_k & 1 & -1 &   \\
     	& -\lambda\beta_{k+1} & 1 & \cdot  \\
     	 & & \cdot & \cdot 
     \end{array}\rb,\qquad k=1,2,\ldots.
     \eeq
     Obviously, $a_\infty=D_1$. Expanding $D_k$ with respect to the first row or the first column, one finds the following 3-term linear recurrence relation
     \beq\label{recdetRCHE}
     D_k=D_{k+1}-\lambda\beta_k D_{k+2}.
     \eeq
     It in turn implies that the ratios $\eta_k=\frac{D_k}{D_{k+1}}$ satisfy a 2-term nonlinear recurrence $\eta_k=1-\lambda\beta_k/\eta_{k+1}$. Since $D_k\to1$ as $k\to\infty$, the solution of this recurrence can be written as
     \beq\label{etak}
     \eta_k=1-\frac{\lambda\beta_k}{1-\frac{\lambda\beta_{k+1}}{1-\ldots}}.
     \eeq
     The statement of the lemma then immediately follows from $\ln D_1=\sum_{k=1}^\infty\ln\eta_k$.
     \epf
     \begin{cor}
     	The integers $N_\mu$ in \eqref{combiN} are given by the following products of binomial coefficients:
     	\beq\label{coefsBC}
     	N_\mu=\frac{2n}{\mu_1}\prod_{m=1}^{\ell-1}{\mu_m+\mu_{m+1}-1 \choose \mu_{m+1}}.
     	\eeq
     \end{cor}
     \pf The factor $\beta_k^{\mu_1}$ can appear in \eqref{combiN} only from the $\mu_1$-th term in the expansion of the logarithm in $\ln\lb 1-\frac{\lambda\beta_k}{1-\frac{\lambda\beta_{k+1}}{1-\ldots}}\rb$, i.e. from $-\frac1{\mu_1}\lb \frac{\lambda\beta_k}{1-\frac{\lambda\beta_{k+1}}{1-\ldots}}\rb^{\mu_1}$. Likewise, the next factor $\beta_{k+1}^{\mu_2}$ can only be produced by the $\mu_2$-th term in the binomial expansion of $\lb 1-\frac{\lambda\beta_{k+1}}{1-\frac{\lambda\beta_{k+2}}{1-\ldots}}\rb^{-\mu_1}$ given by
     $\lb-1\rb^{\mu_2}{-\mu_1 \choose \mu_2}\lb\frac{\lambda\beta_{k+1}}{1-\frac{\lambda\beta_{k+2}}{1-\ldots}}\rb^{\mu_2}$. Continuing the same procedure, we see that the coefficient of $\beta_{k}^{\mu_1}\beta_{k+1}^{\mu_2}\ldots\beta_{k+\ell}^{\mu_\ell}$ in the expansion of $\ln\eta_k$ is equal to
     \beq
     -\lb-1\rb^{\mu_2+\mu_3+\ldots+\mu_\ell}\frac{1}{\mu_1}{-\mu_1 \choose \mu_2}
     {-\mu_2 \choose \mu_3}\ldots {-\mu_{\ell-1} \choose \mu_\ell}.
     \eeq 
     The representation \eqref{coefsBC} is nothing but a rewrite of the latter expression taking into account the coefficient $\frac1{2n}$ in front of $\operatorname{Tr} A^{2n}$ in \eqref{lndet}. \epf
     
     The formulas such as \eqref{lndet}, \eqref{combiN}, \eqref{coefsBC}  allow to compute the expansion of $\ln a_\infty$ in powers of $\lambda$ to arbitrary order. In practice, it is more convenient to use \eqref{contfrRCHE}, truncate the infinite fraction at appropriate order and expand the result. The coefficients $\mathsf c_n$ of
     \beq
     \ln a_\infty=\sum_{n=1}^\infty \mathsf{c}_n\lambda^n,
     \eeq
     are then given by linear combinations of sums $\sum_{k=1}^\infty \beta_{k}^{\mu_1}\beta_{k+1}^{\mu_2}\ldots\beta_{k+\ell}^{\mu_\ell}$, cf \eqref{expltraces}. Since $\beta_k$ defined by \eqref{betaRCHE} is a rational function of $k$, such sums can be calculated in a closed form in terms of rational and polygamma functions. Indeed, using partial fraction decomposition of $\beta_{k}^{\mu_1}\beta_{k+1}^{\mu_2}\ldots\beta_{k+\ell}^{\mu_\ell}$ with respect to $k$, any such sum can be written as a linear combination of sums of the form
     \begin{subequations}
     \begin{alignat}{4}
     \sum_{k=1}^\infty\frac{1}{\lb k-x\rb^{n+1}}&=\lb-1\rb^{n+1}\frac{\psi^{(n)}\lb 1-x\rb}{n!},&&n\ge1,\\
     \sum_{k=1}^\infty\lb \sum_{m=1}^n\frac{y_m}{ k-x_m}\rb&=-\sum_{m=1}^ny_m\psi\lb 1-x_m\rb,\qquad\qquad
     &&\sum_{m=1}^ny_m=0.
     \end{alignat}
     \end{subequations}
     The property $\sum_{m=1}^n y_m=0$ is ensured automatically by the asymptotic behavior $\beta_k=O\lb\frac1{k^2}\rb$ as $k\to\infty$.  
         For example, it follows from \eqref{expltracesa} that
     \begin{subequations}
     \begin{align}
     \mathsf c_1=&\,-\frac{\frac14-\theta_0^2+\theta_1^2-\omega^2}{4\omega\lb\frac14-\omega^2\rb}
     \,\Bigl(\psi\lb\tfrac12-\theta_0+\theta_1+\omega\rb-\psi\lb\tfrac12-\theta_0+\theta_1-\omega\rb\Bigr)+\frac{\theta_0+\theta_1}{2\lb \frac14-\omega^2\rb},\\
     \mathsf c_2=&\,-\frac{\lb\frac14-\theta_0^2+\theta_1^2-\omega^2\rb^2}{32\omega^2\lb\frac14-\omega^2\rb^2}\Bigl(\psi^{(1)}\lb\tfrac12-\theta_0+\theta_1+\omega\rb+\psi^{(1)}\lb\tfrac12-\theta_0+\theta_1-\omega\rb\Bigr)+\\
     \nonumber
     &\,+\lb\frac{\lb 60\omega^4-35\omega^2+2\rb\lb\theta_0^2-\theta_1^2\rb^2}{256\omega^3\lb\frac14-\omega^2\rb^3\lb 1-\omega^2\rb}-\frac{3\lb\theta_0^2+\theta_1^2\rb}{32\omega\lb1-\omega^2\rb\lb\frac14-\omega^2\rb}-\frac{\lb1-12\omega^2\rb\lb\theta_0^2-\theta_1^2\rb}{64\omega^3\lb\frac14-\omega^2\rb^2}+\frac{2-3\omega^2}{64\omega^3\lb 1-\omega^2\rb}\rb\times\\
     \nonumber &\,\qquad\times \Bigl(\psi\lb\tfrac12-\theta_0+\theta_1+\omega\rb-\psi\lb\tfrac12-\theta_0+\theta_1-\omega\rb\Bigr)+\\
     \nonumber &\,+\frac{\theta_0+\theta_1}{4\lb\frac14-\omega^2\rb^2}-\frac{3\lb\theta_0-\theta_1\rb}{32\lb\frac14-\omega^2\rb\lb 1-\omega^2\rb}-\frac{\lb25-52\omega^2\rb\lb\theta_0-\theta_1\rb\lb\theta_0+\theta_1\rb^2}{128\lb\frac14-\omega^2\rb^3\lb 1-\omega^2\rb}.
     \end{align}
     \end{subequations}
     This calculation confirms the validity of the Trieste formula for RCHE \cite{BILT22}. We have compared our rigorous results against CFT predictions going up to corrections of order $\lambda^5$ and found complete agreement. There are no obstacles other than computer time to check higher orders which essentially amounts to comparison of huge rational expressions.

     \subsection{Heun equation (HE)}
     In the case of the usual Heun equation \eqref{usheun}, it is convenient to consider a modification of the Schäfke-Schmidt formula. Instead of $u\lb z\rb$ defined by \eqref{defuz}, we introduce
     \beq\label{defuzheun}
     u\lb z\rb:=z^{-\frac12+\theta_0}\lb 1-z\rb^{-\frac12-\theta_1}\lb t-z\rb^{-\frac12+\theta_t}
     \psi_+^{[0]}\lb z\rb=t^{-\frac12+\theta_t}\Bigl[ 1+\sum_{k=1}^\infty u_kz^k\Bigr].
     \eeq
     The effect of the extra factor $\lb t-z\rb^{-\frac12+\theta_t}$ in the above  is a slight change of the formula for the connection coefficient as compared to \eqref{SSfor}:
     \beq
     \mathsf C\lb\theta_0,\theta_1\rb=\Gamma\lb 2\theta_1\rb \lb 1-\tfrac1t\rb^{\frac12-\theta_t}\lim_{k\to\infty}k^{1-2\theta_1}u_k.
     \eeq
     The function $u\lb z\rb$ defined by \eqref{defuzheun} satisfies the canonical form of HE,
     \beq\label{canheun}
     \left[\frac{d^2}{dz^2}+\lb\frac{1-2\theta_0}{z}+\frac{1+2\theta_1}{z-1}+\frac{1-2\theta_t}{z-t}\rb\frac{d}{dz}+\frac{\lb\lb\theta_0-\theta_1+\theta_t-1\rb^2-\theta_\infty^2 \rb z-q}{z\lb z-1\rb\lb z-t\rb}\right]u\lb z\rb=0,
     \eeq
     where
     \beq
     q=t\lb \theta_0-\theta_1-\tfrac12\rb^2-t \omega^2+\lb\theta_0+\theta_t-\tfrac12\rb^2-\theta_0^2-\theta_\infty^2+\omega^2.
     \eeq
     
     The main advantage of \eqref{canheun}, obtained at the expense of loosing the symmetry $\theta_t\leftrightarrow -\theta_t$, is that it yields a 3-term recurrence relation for the coefficients $u_k$. Also note that for $t\to\infty$, equation \eqref{canheun} transforms into hypergeometric equation \eqref{hypercan} (with $\theta_\infty$ replaced by $\omega$), which implies that
    \beq
u_k^{t\to\infty}=\frac{\lb \tfrac12-\theta_0+\theta_1+\omega\rb_k \lb \tfrac12-\theta_0+\theta_1-\omega\rb_k}{k!\,\lb 1-2\theta_0\rb_k}.
\eeq  
     Introducing the rescaled coefficients $a_k=u_k/u_k^{t\to\infty}$, it becomes straightforward to write the Heun counterpart of Lemma~\eqref{lemmaRCHE1}: 
     \begin{lemma}
     	The connection matrix relating normalized Frobenius solutions of the HE at $z=0,1$ is given by 
     	\beq
     	\mathsf C\lb\theta_0,\theta_1\rb=\frac{\Gamma\lb 1-2  \theta_0\rb\Gamma\lb2\theta_1\rb}{\Gamma\lb\tfrac12+\theta_1-\theta_0+\omega\rb \Gamma\lb\tfrac12+\theta_1-\theta_0-\omega\rb}\,\lb 1-\lambda\rb^{\frac12-\theta_t} a_\infty,
     	\eeq
    where  $\lambda=\frac1t$ and $a_\infty=\lim\limits_{k\to\infty}a_k$ is the limit of the sequence defined by the 3-term recurrence relation
    \beq\label{3termHE}
    a_{k+1}-a_k=-\lambda \lb\alpha_k a_k+\beta_k a_{k-1}\rb,
    \eeq
    with $a_{-1}=0$, $a_0=1$ and $\alpha_k,\beta_k$ given by \eqref{alphabetaHE}.
     \end{lemma}  
   The coefficients $\alpha_k$, $\beta_k$ in \eqref{alphabetaHE} no longer tend to $0$ as $k\to\infty$. This makes infinite determinants generalizing \eqref{infdetRCHE} ill-defined. Nevertheless it is still possible to obtain an analog of Lemma~\ref{contfrRCHElem}.
   \begin{lemma}
   	\label{contfrHElem}
   	The quantity $a_\infty$ admits the following infinite fraction representation:
   	\beq\label{contfrHE}
   	\ln a_\infty=-\ln\lb1-\lambda\rb+\sum_{k=1}^\infty\ln\lb1-\lambda\alpha_{k-1}-\frac{\lambda\beta_k}{1-\lambda\alpha_{k}-\frac{\lambda\beta_{k+1}}{1-\ldots}}\rb.
   	\eeq
   	\end{lemma}
   	\pf Let $a^{(N)}_k$ denote the solution of the recurrence relation \eqref{3termHE} satisfying the initial conditions $a_{N-1}=0$, $a_{N}=1$. We are ultimately interested in the limiting value of $a^{(0)}_k$ as $k\to\infty$. Since the sequences $a^{(N)}_k$ and $a^{(N+1)}_k$ are linearly independent, there exist $C_1$ and $C_2$ such that $a^{(N+2)}_k=C_1a^{(N+1)}_k+C_2a^{(N)}_k$. It can be deduced from the initial conditions on three sequences that $C_1=t\beta_{N+1}^{-1}-\alpha_N\beta_{N+1}^{-1}$, $C_2=-t\beta_{N+1}^{-1}$, so that
   	\beq\label{dualrec}
   	a^{(N)}_k-a^{(N+1)}_k=-\lambda\lb \alpha_N a^{(N+1)}_k+\beta_{N+1}a^{(N+2)}_k\rb.
   	\eeq

   	Define $D_{N+1}:=\lim\limits_{k\to\infty}a^{(N)}_k$. The existence of the limit is ensured by the Schäfke-Schmidt formula. Note in particular that the quantity $a_\infty$ that we are after coincides with $D_1$. The recurrence \eqref{dualrec} implies that
   	\beq
   	D_{N}-D_{N+1}=-\lambda\lb \alpha_{N-1} D_{N+1}+\beta_{N}D_{N+2}\rb.
   	\eeq
   	This relation is the Heun counterpart of \eqref{recdetRCHE}. Now we can proceed similarly to the proof of Lemma~\ref{contfrRCHElem}. Introducing the ratios $\eta_k=\frac{D_k}{D_{k+1}}$, one obtains a 2-term Riccati recurrence  $\eta_k=1-\lambda\alpha_{k-1}-\lambda\beta_k/\eta_{k+1}$. It
    is solved by the continued fraction
   	\beq
   	\eta_k=1-\lambda\alpha_{k-1}-\frac{\lambda\beta_k}{1-\lambda\alpha_{k}-\frac{\lambda\beta_{k+1}}{1-\ldots}}.
   	\eeq
   	We may write
   	\beq\label{DNp1}
   	D_1=\ln D_{N+1}+\sum_{k=1}^{N}\ln \eta_k .
   	\eeq
   	The existence of the  limit $N\to\infty$ of the sum in the second term does not  require that $\alpha_k,\beta_k\to0$ as $k\to\infty$. In fact, it suffices to have ${\alpha_{k-1}+\beta_k=O\lb \frac1{k^2}\rb}$, which holds for $\alpha_k,\beta_k$ defined by \eqref{alphabetaHE}. Therefore it remains to compute the limit
   	\beq
   	D_\infty=\lim_{N\to\infty}D_{N+1}=\lim_{N\to\infty}\lim_{M\to\infty}
   	\operatorname{det}\lb\begin{array}{ccccc}1-\lambda\alpha_N & -1 &  & & \\
   		-\lambda\beta_{N+1} & 1-\lambda\alpha_{N+1} & -1 &  & \\
   		& -\lambda\beta_{N+2} & \cdot & \cdot  &\\
   		& & \cdot & \cdot & -1 \\
   		& & & -\lambda\beta_{M+N} & 1-\lambda\alpha_{M+N}
   	\end{array}\rb.
   	\eeq
   	Since $\alpha_k=-1+O\lb\tfrac 1k\rb$, $\beta_k=1+O\lb\tfrac 1k\rb$ as $k\to\infty$, it follows that $D_\infty$ coincides with the limiting value $\lim\limits_{M\to\infty} \tilde D_M$ of the determinant of an $M\times M$ tridiagonal Toeplitz matrix, 
   	\beq
   	\tilde D_M=\operatorname{det}\lb\begin{array}{cccc}
   1+\lambda & -1 & & \\
   -\lambda & 1+\lambda & \cdot &  \\
   & \cdot &\cdot & -1 \\
   & & -\lambda & 1+\lambda	
   \end{array}\rb.
   	\eeq
   	The latter determinant can be easily evaluated (e.g. by induction on $M$) to $\tilde D_M=\sum_{k=0}^{M}\lambda^k$, which implies that $D_\infty=1/\lb 1-\lambda\rb$. In combination with \eqref{DNp1}, this yields \eqref{contfrHE} and concludes the proof of Theorem~\ref{thmB}.
   	\epf
   	
   	As before for RCHE, the coefficients in the expansion $\ln a_\infty=\sum_{n=1}^\infty\mathsf c_n\lambda^n$ are thus given by sums of explicit rational expressions and can be evaluated in a closed form. For example,
   	\begin{align}
   	\mathsf c_1=&\,1-\sum_{k=1}^\infty\lb\alpha_{k-1}+\beta_k\rb=\tfrac12-\theta_t+\mathsf f_1,
   	\end{align}
   	where $\mathsf f_1$ is given by \eqref{coeff1}. Similarly computing subsequent coefficients, we have successfully checked the Trieste formula \eqref{triestef} up to order $\lambda^3$. 
          
     \subsection{Confluent Heun equation (CHE)}
     The case of the CHE 
     \beq
     \left[\frac{d^2}{dz^2}+\frac{\frac14-\theta_0^2}{z^2}+\frac{\frac14-\theta_1^2}{\lb z-1\rb^2}-\frac{\lambda^2}{4}-\frac{\lambda\theta_*}{z}+\frac{\theta_0^2+\theta_1^2 -\omega^2-\frac14}{z\lb z-1\rb}\right]\psi\lb z\rb=0,
     \eeq
     is intermediate between the RCHE and HE. Since it can be analyzed in a similar manner, we limit ourselves to the statement of the final result.
     \begin{customthm}{C}\label{thmC}
     The connection relation between the normalized Frobenius solutions $\psi^{[0]}_{\pm}\lb z\rb$, $\psi^{[1]}_{\pm}\lb z\rb$ of the CHE is given by \eqref{connnn}, with $\mathsf C\lb\theta_0,\theta_1\rb$ defined by
         \beq\label{CoCHE}
     \mathsf C\lb\theta_0,\theta_1\rb=\frac{\Gamma\lb 1-2  \theta_0\rb\Gamma\lb2\theta_1\rb\, e^{\lambda/2}}{\Gamma\lb\tfrac12+\theta_1-\theta_0+\omega\rb \Gamma\lb\tfrac12+\theta_1-\theta_0-\omega\rb}\,\exp \sum_{k=1}^\infty\ln\lb1-\lambda\alpha_{k-1}-\frac{\lambda\beta_k}{1-\lambda\alpha_{k}-\frac{\lambda\beta_{k+1}}{1-\ldots}}\rb,
     \eeq
     where
     \begin{subequations}\label{alphabetaCHE}
     	\begin{align}
     	\label{alphaCHE}
     	\alpha_k=&\, \frac{k+\frac12-\theta_0-\theta_*}{\lb k+\frac12-\theta_0+\theta_1\rb^2-\omega^2},\\
     	\label{betaCHE}
     	\beta_k=&\,-\frac{k\lb k-2\theta_0\rb
     		\lb k-\theta_0+\theta_1-\theta_*\rb}{\lb \lb k+\frac12-\theta_0+\theta_1\rb^2-\omega^2\rb
     		\lb \lb k-\frac12-\theta_0+\theta_1\rb^2-\omega^2\rb}.
     	\end{align} 	
     \end{subequations}
     \end{customthm}
    
     At the formal level, the formulas \eqref{CoCHE}--\eqref{alphabetaCHE} can be obtained from their Heun analogs \eqref{CoHE}--\eqref{alphabetaHE} in Theorem~\ref{thmB} in the limit
     \beq
     \theta_t=\frac{\Lambda+\theta_*}{2},\qquad 
     \theta_\infty=\frac{\Lambda-\theta_*}{2},\qquad
     \lambda_{\text{HE}}=\frac{\lambda_{\text{CHE}}}{\Lambda},\qquad \Lambda\to\infty.
     \eeq
     Further limit to RCHE is more subtle and involves in addition a $\lambda$-dependent redefinition of the accessory parameter $\omega$.

        \section{Discussion and outlook}
        We have developed a procedure of the perturbative solution of the connection problem for the usual, confluent and reduced confluent Heun equation between Frobenius solutions associated to different Fuchsian singular points. It confirms the validity of the Trieste formula \eqref{triestef} and its confluent variants.
        Compared to CFT approach, Theorems~\ref{thmA}, \ref{thmB}, \ref{thmC} become particularly efficient  when monodromy exponents $\boldsymbol{\theta}$ and accessory parameter $\omega$ are assigned specific numerical values, since in the former setting one needs to keep $\theta_0,\theta_1$ arbitrary for the computation of derivatives $\frac{\partial\mathcal W}{\partial\theta_0}$, $\frac{\partial\mathcal W}{\partial\theta_1}$.
        
        It would be interesting to generalize our approach to the connection problem involving irregular singularities using an appropriate modification of the Schäfke-Schmidt formula. We note that in this case there also exist perturbative expansions \cite{BILT22} predicted by CFT. In addition to CHE and RCHE, this concerns the doubly-confluent Heun equation and its reduced and doubly-reduced version.
        
        Another appealing direction would be to study the connection problem for different confluent Heun equations at strong coupling $|\lambda|\gg 1$ (instead of the weak coupling considered here). Indeed, in this regime there already exist analogs of the Zamolodchikov conjecture \cite{LN} for the confluent and biconfluent Heun equation. Moreover, the expansion of the relevant accessory parameter function in CHE can be obtained from 3-term recurrence relations and continued fractions \cite{Cunha}.
        
        To the authors' knowledge, there is no rigorous proof of the Zamolodchikov relation $\mathcal{E}=t\frac{\partial\mathcal W}{\partial t}$  between the Heun accessory parameter and quasiclassical conformal block available yet. Moreover, without control of analytic properties of such conformal blocks it is unclear how to interpret this statement beyond the level of formal series. The Trieste formula, on the other hand, can be formulated in purely mathematical terms. One may \textit{define} the function $\mathcal W\lb t\rb$ by \eqref{accmon} in which case \eqref{triestef} yields a highly nontrivial conjectural relation between the Heun accessory parameter of Floquet type and the connection matrix. We expect it to be related to the extended symplectic structure on the space of monodromy data of Fuchsian systems introduced in \cite{BK}.

\end{document}